\documentclass[aps,prl,twocolumn,showpacs,groupedaddress,preprintnumbers]{revtex4}
\usepackage{graphicx}
\usepackage{dcolumn}
\usepackage{bm}
\begin{document}
\preprint{astro-ph/0904.2371}

\title{Cosmic Ray Proton Background Could Explain ATIC Electron Excess}
\affiliation{Southern University, Baton Rouge, LA 70813}
\author{A.R. Fazely}\email{ali_fazely@subr.edu}\affiliation{Southern University, Baton Rouge, LA 70813}
\author{R.M. Gunasingha}\altaffiliation{Present address: Box 3155, Duke University Medical Center, Durham, NC 27710}\affiliation{Southern University, Baton Rouge, LA 70813}
\author{S.V. Ter-Antonyan}\affiliation{Southern University, Baton Rouge, LA 70813}
\begin{abstract}
We show that the excess in the Galactic electron flux
recently
published by Chang, et al. (Nature, 20 Nov. 2008) can have a simple methodical origin 
due to
a contribution from misidentified proton induced electron-like events in the 
ATIC detector.
A subtraction of the cosmic ray proton component from the published ATIC 
electron flux eliminates this excess in the range of 300 to 800 GeV.
\end{abstract}


\maketitle
An excess in the Galactic electron flux in the range of $300-800$ GeV
recently published by J. Chang, et al, \cite{chang}
has led to numerous speculations \cite{chang,zhang,ilia,koji,hong,nir,stef} 
about the origin of
the purported excess such as annihilation of dark matter (DM) 
\cite{chang,zhang},
decaying DM \cite{zhang,ilia}, decay of lightest superparticle DM \cite{koji},
interaction of high energy cosmic rays with
photon background near accelerating sites \cite{hong}, a few nearby
SNR \cite{nir}, and distant young pulsars \cite{stef}. All theoretical 
predictions
\cite{chang,zhang,ilia,koji,hong,nir,stef} describe with more or less 
efficiency the published energy
spectrum reported by Chang et al.,\cite{chang} in the energy region of 
$300-800$ GeV depending on a number of applied free model parameters.

Chang, et al.,\cite{chang} use an excess of 70 events, spread over an energy range 
of 500 GeV, to 
report a $\sim6\sigma$ signal based on a 210-event sample, 
where a background of 140 events was estimated with a 
GALPROP \cite{GALPROP} calculation,  
neglecting both systematic uncertainties  
associated with the GALPROP prediction  \cite{GALPROP} as well as those of 
the ATIC instrument.
Not noted in reference \cite{chang}, the expected fluctuation of the Galactic 
electron flux due to the stochastic nature of sources \cite{Swordy,Moskal} 
alone could introduce more than a $20\%$ uncertainty for energies above 
$300$ GeV. 
A major source
of background of the ATIC electron experimental data is
due to the primary cosmic ray proton component where the bulk of the proton
energy is transferred to a single particle, for example, a $\pi^0$ in the 
carbon-scintillator target of the ATIC detector.

In this letter, following the principle of Occam's Razor mentioned in the 
title of a recent paper by Stefano~Profumo \cite{stef}, we investigated 
the methodical origin of the reported excess.
We studied the contributions due to high energy primary protons
and atmospheric $e^{\pm}$ and $\gamma$ to the reported Galactic
electron flux in the ATIC data. 
The primary proton and atmospheric $e^{\pm}$ and $\gamma$-background
components at the level
of ATIC flights ($\sim5$g/cm$^2$) \cite{chang} and energies more than $50$ 
GeV have an energy spectral power index $\gamma_A\simeq-2.7$ and could change
significantly the observed spectral index $\gamma_e\simeq-3.2$ of Galactic 
electrons, especially above the TeV energy region, where an exponential
energy cut-off is expected \cite{HESS}.

ATIC, the acronym for Advanced Thin Ionization Calorimeter
was designed for multiple, long duration balloon flights,
to measure nucleonic cosmic ray spectra from
10 GeV to near 100 TeV total energy, using a fully active Bismuth
Germanate (BGO) calorimeter. It was equipped with the first large
area mosaic of small fully depleted silicon detector wafers for
charge identification of cosmic rays from H to Fe. As a particle tracking
system, three projective layers of x-y scintillator hodoscopes were
target\cite{guzik}.
The experiment had no magnetic field and was designed for cosmic ray 
elemental composition measurements with
limited particle ID (PID)
capability especially for the types of measurements mentioned in 
reference \cite{chang}. 

In this letter, we used the published data of reference \cite{chang} and 
applied their cut parameters, without delving into a detailed analysis of 
their validity, to a GEANT Monte Carlo (MC) \cite{GEANT}.    
The primary energy nuclei spectra were taken from power law approximations of
balloon and satellite data \cite{PB} 
\begin{equation}
\frac{d\Im_A}{dE_A}=\Phi_{A}E_A^{-\gamma_A},
\end{equation}
where $\Phi_A$ and $\gamma_A$ are the scale parameters
and spectral indices for $A\equiv p,He,O$-like, and 
$Fe$-like nuclear species \cite{PB}.

The flux of the $p,e^{\pm}$ and $\gamma$-background components of
cosmic rays at the ATIC
level were computed using CORSIKA shower simulation code \cite{CORSIKA} in the
frames of SIBYLL \cite{SIBYLL} interaction model, taking into account the
South Pole atmosphere and magnetic field. The estimated contributions of the
secondary atmospheric $\gamma$ and $e^{\pm}$-components to the Galactic 
electron flux varied slightly from the corresponding ATIC 
results of reference \cite{chang}, 
where the proton background was said to be included. This result has indicated
that proton-induced e-like events were substantially underestimated in reference 
\cite{chang}. Because 
the contribution of $e^{\pm}$ and $\gamma$ atmospheric components are negligible to the overall data, 
we have omitted the 
details here and instead concentrate on the main background which comes from
cosmic ray protons.

The contribution of the primary proton background to the 
Galactic electron flux was calculated using the proton 
spectra at the ATIC level
in a GEANT-3.21 calculation with the GCALOR routines in the frame of the FLUKA 
interaction model for the ATIC detector \cite{gab}.  
This calculation takes into account the ATIC geometry with the
silicon matrix, scintillator hodoscopes, graphite targets and eight 
layers of x-y
BGO crystals, containing appropriate material compositions \cite{guzik}.
The total number of simulated proton and electron events were  
$100,000$ and $10,000$, respectively.

\begin{figure} 
\begin{center}
\includegraphics[width=9cm, height=9cm]{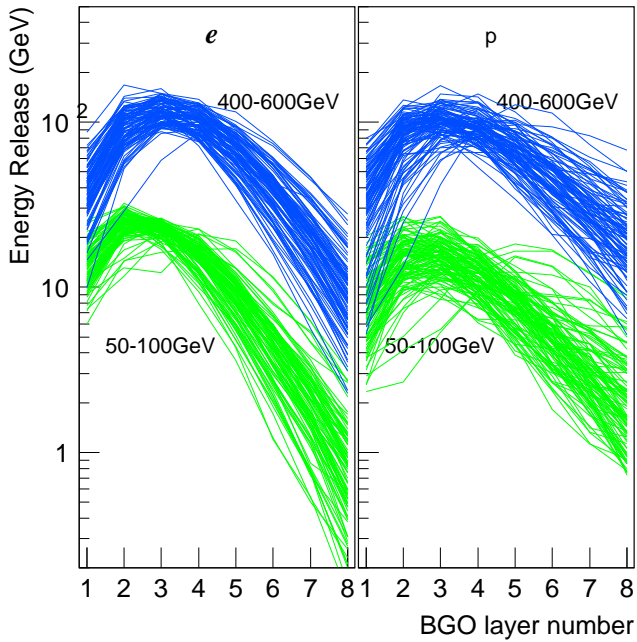}
\end{center}
\caption{GEANT simulated 
electron (left panel) and proton (right panel) for 100 cascade events showing the 
longitudinal profiles 
in the ATIC BGO calorimeter at 
energies of $400 - 600$ GeV (upper lines) and at $50-100$ GeV (lower lines).}
\label{cas}
\end{figure}

These GCALOR calculations for the ATIC detector were previously compared 
with a FLUKA-2006 calculation \cite{guna} as well as with those of an 
ATIC test-beam run at CERN \cite{asif}. Furthermore, we also ran FLUKA-2006
for the ATIC configuration and results were consistent with those of the 
GCALOR calculations. Figure \ref{cas} shows 
the electron (left panel) and proton (right panel) induced cascade profiles
in the ATIC BGO calorimeter after $e$-like cascade selection criteria:
$n_{\max}\leq5$, $E_{d,\max}>E_{d,\max+1}>E_{d,\max+2}\dots E_{d,8}$,
$E_{d,1}>2$ GeV and $E_{d,8}>25$ MeV,
where $E_{d,n}$ is the detected energy in the $n$-th BGO layers
and $E_{d,\max}=\max{(E_{d,n})}$ for $n=1,\dots8$. The lower and upper lines
in Figure \ref{cas} correspond to $50-100$ GeV and $400-600$ GeV detected
energies respectively. The selection criteria  rejects $74\%$ of proton events 
with visible energies $E_v>50$ GeV without significant change of
the electron cascade detection efficiency. This result is in close
agreement with the $e$-cascade selection efficiency
($\sim90\%$) and proton rejection efficiency ($\sim75\%$) 
of reference \cite{chang}.

\begin{figure} 
\begin{center}
\includegraphics[width=9cm, height=9cm]{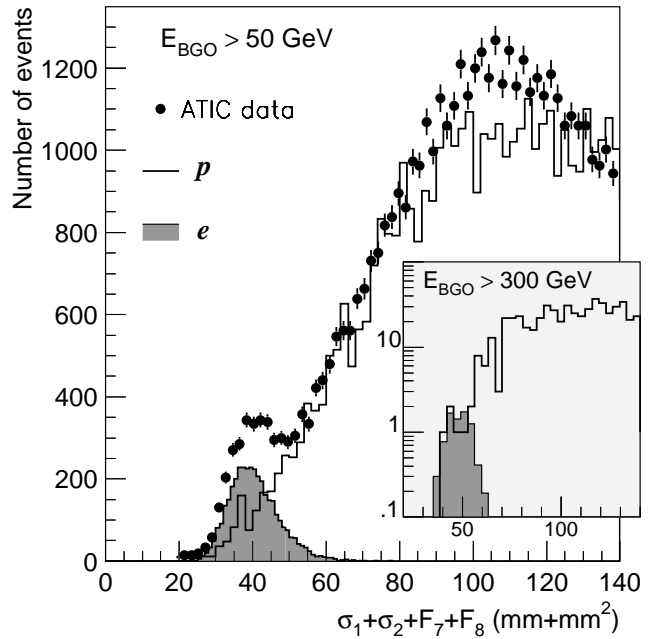}
\end{center}
\caption{Distribution of combined cascade parameter
$\sigma_1+\sigma_2+F_1+F_2$ for
detected energies $E_{BGO}=E_d>50$ GeV 
(and $300$ GeV, inset histograms). 
The symbols are the ATIC data from \cite{chang}. The large (line)
and shaded histograms are the expected distributions
for primary proton and electron passing through the BGO 
calorimeter.}
\label{atic_geant}
\end{figure}

We simulated the lateral distribution of the cascade parameter
$F\equiv\sigma_1+\sigma_2+F_7+F_8$ employed in reference \cite{chang} for 
proton induced $e$-like cascades and pure $e$-cascades. Figure \ref{atic_geant}
shows the published data of reference \cite{chang} and our calculations.  
Note the indices of $\sigma$ and $F$ refer to the
BGO layers. Our simulated data for $E_v>50$ GeV 
was normalized to the corresponding statistics of the ATIC data \cite{chang} 
taking 
into account energy-dependent ratio $(d\Im_p/dE_p)/(d\Im_e/dE_e)$ 
of the expected proton flux from expression (1) and the expected electron 
flux $d\Im_e/dE_e=\alpha E_e^{-3.05}\exp(-E_e/E_0)$ from the HESS experiment 
\cite{HESS}. The results of the GEANT simulations for visible energies 
$E_v>300$ GeV are shown in the inset histogram, as well.

\begin{figure} 
\begin{center}
\includegraphics[width=9cm, height=9cm]{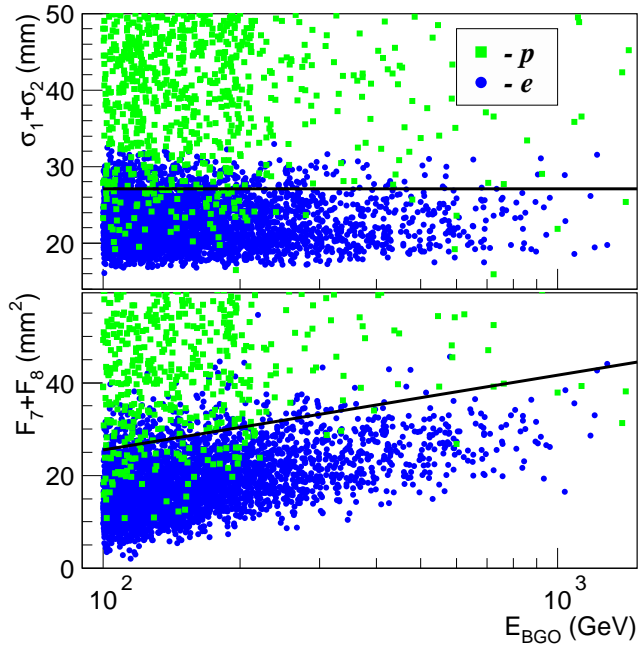}
\end{center}
\caption{2-dimensional $((\sigma_1+\sigma_2),E_{BGO})$, (upper panel)
and $((F_7+F_8),E_{BGO})$, (lower panel) scatter plots for 5000 simulated
proton and electron events in the ATIC BGO calorimeter. The lines
are the corresponding proton rejecting cuts.}
\label{cuts}
\end{figure}

It is observed that the contribution of primary
proton induced $e$-like cascades in the ATIC BGO calorimeter to the flux
of Galactic electrons is quite substantial in the region of the reported
excess ($E_d>300$ GeV).
\cite{chang}. 
"Pure" e-cascade pattern recognition
in the ATIC data is performed
using diagonal energy dependent cuts \cite{chang} in a 2-dimensional (2d)
($\sigma_1+\sigma_2$, $F_1+F_2$) space. The corresponding 
GEANT simulated $((\sigma_1+\sigma_2),E_{d})$ and $((F_7+F_8),E_{d})$
scatter plots are presented in Figure \ref{cuts}.
The number of simulated events was taken to
be equal at 5000 events for both electrons and protons with $E_d>50$ GeV. 
Note that the actual ratio of $p/e$ in cosmic rays is approximately 300-500
for these energies.

\begin{figure}[h] 
\begin{center}
\includegraphics[width=9cm, height=9cm]{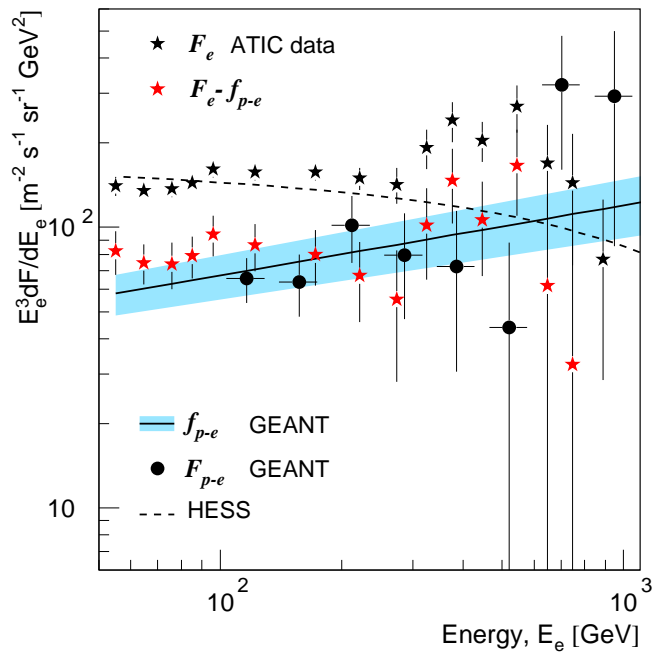}
\end{center}
\caption{Galactic electron and misidentified $e$-like energy spectra.
The solid black stars ($F_e$) are the data of reference \cite{chang}, the 
red stars
are the ATIC data after subtraction ($F_e-f_{p-e}$) of $e$-like proton
energy spectrum ($f_{p-e}$,
the solid line with blue shaded area of errors)
obtained from the corresponding integral flux, the solid black circles are
the GEANT simulated proton induced $e$-like events energy spectrum.  
The dashed line is the power law energy spectrum with exponential 
cut-off from HESS \cite{HESS}.}
\label{ee}
\end{figure}

Because the 2d selection criteria (cuts) applied in the 
published data were unavailable
from the literature, we derived these 2d cuts (lines in Figure \ref{cuts}) from
the GEANT simulated database using known 
$84\%$ efficiency of e-cascade selection from \cite{chang}.
Lines shown in Figure \ref{cuts} represent cuts corresponding to
the same afore-mentioned selection efficiency of e-cascade from GEANT 
simulated database.\\
The energy spectrum of proton induced and subsequent       
misidentified $e$-like events expected in the ATIC data
were calculated according to the expression
\begin{equation}
F_{p-e}=\frac{\Im_p}{N_{tot}}\frac{\Delta N}{\Delta E\cdot\delta_e\cdot\xi}
\end{equation}
where $E_e=E_d/\varepsilon$ and $\varepsilon$ is the average fraction 
of released energy due to e-cascade passing through the BGO calorimeter \cite{chang},
$\Im_p$ is the integral energy spectrum of the primary protons from 
expression (1) for 
$E_{p,\min}=100$ GeV, $N_{tot}$ is the total number of simulated proton events,
$\Delta N$ and $\Delta E$ are the number of selected
e-like proton induced events in the given $\Delta E$ energy bin, $\delta_e=0.84$ 
\cite{chang} is the efficiency of 2-dimensional e-cascade selection cuts, and
$\xi(E_e)\simeq\xi\equiv\Im_{e,ATIC}/\Im_{e,Prim}\simeq0.78$ 
is the atmospheric reduction factor taken from reference \cite{chang1}.
The corresponding results are presented in Figure \ref{ee} (black circles) 
in comparison with published data of reference \cite{chang} 
(black star symbols). The accuracy of the obtained 
energy spectrum was improved using the derivative of the corresponding  
integral spectra due to {\em{a priori}} known 
spectral index ($\gamma_p=-2.75$) of proton induced $e$-like cascades.
The corresponding
differential energy spectrum, normalized for energy $E_e = 100$ GeV, is shown in 
Figure \ref{ee} (black line with blue-shaded area for the expected errors), as well. 
Red star
symbols are the residual energy spectrum of Galactic electrons after
subtraction of the expected proton induced $e$-like cascade from the
reported data of reference \cite{chang}.\\
Estimated contribution of the expected e-like proton background 
is turned out to be  $96 \pm 9.8$ as compared to the reported 70-event 
excess of reference \cite{chang}, thus rejecting the conclusion of exotic 
electrons.\\  
In summary, Figure \ref{ee} clearly shows that the MC proton spectrum has a 
substantial 
contribution in the energy region of more that 300 GeV where an excess
of e-like events was reported \cite{chang}. Our analysis indicates that 
a calorimeter such as ATIC with its limited PID capability could easily 
misidentify protons for electrons and thereby yield unwarranted physics 
conclusions. The estimate background of $96 \pm 9.8$ events in the 
energy region of 300 - 800 GeV completely explains the 70-event
excess of reference \cite{chang}. We therefore conclude that the contribution
from cosmic ray protons at energies above 300 GeV has not been properly 
included in reference \cite{chang}. This
analysis shows that the results of reference \cite{chang} does not merit any 
exotic conclusion and that the electron spectrum is consistent 
with that reported by the AMS collaboration \cite{AMS}. 

The authors gratefully acknowledge a grant from NASA
under the contract No. NNX07AE49G.

\end{document}